\documentclass[epsfig,amssymb]{article}
\usepackage{epsfig}
\newcommand{\beq}{\begin{equation}}
\newcommand{\eeq}{\end{equation}}
\newcommand{\be}{\begin{equation}}
\newcommand{\ee}{\end{equation}}
\newcommand{\beqa}{\begin{eqnarray}}
\newcommand{\eeqa}{\end{eqnarray}}
\newcommand{\oh}{\frac{1}{2}}

\newcommand{\ZP}{Z. Phys. }

\title{\bf Resonance Conversion as a Catalyser of Nuclear Reaction{\bf s}$^{*}$}
\author{
KARPESHIN  Feodor$^{1,2}$,
ZHANG Jingbo$^{1}$, and
ZHANG Weining$^{1}$\\
{\small\sl
$^{1}$Science Center and Department of Physics, Harbin Institute of
Technology, Harbin 150001}\\
{\small\sl
$^{2}$Fock Institute of Physics, St. Petersburg State University, RU-198504
St. Petersburg, Russia}\\
}
\date{}

\begin{document}

\baselineskip 13pt
\twocolumn[
\vspace{-1.5cm}
\maketitle
\vspace{-0.5cm}
\begin{center}
\begin{minipage}{162mm}
{\sl{

     It is shown that resonance interal conversion 
 offers a
feasible tool for mastering nuclear processes with laser or synchrotron
radiation.
Physics of the process is discussed in detail in historical asprct.
Possible way of experimental
applicaytion is shown in the case of the $M1$ 70.6-keV transition
in nuclei of $^{169}$Yb. Nuclear transition rate in hydrogenlike
ions of this nuclide can be enhanced by up to four orders of magnitude. \\[1ex]

}}
\end{minipage}
\end{center}
]

\footnotetext{\hspace{-0.5cm}$^*$Supported by
Russian Foundation for
Basic Research(05-02-17340) and
Scientific Research Foundation of
Harbin Institute of Technology (HIT.2003.23)}

\section{Introduction}

In 1939, Bohr and Wheeler proposed the version of nuclear theory,
which since that time is applied for description of this wonderful process
\cite{Bwhe}.
Ten years later, Wheeler told out an idea that fission of $^{238}{\rm U}$
nucleus in the muonic atom can be induced by a radiationless muon
transition $2s \rightarrow 1s$[2].
Later investigation showed that the population of the $2s$ level
is less probable, of a few percent of that for the $2p$ state,
and additionally, the radiative $2s \rightarrow 2p$ transition
makes a strong competition. However, it was shown that the radiationless
transition probability for higher multipoles, $E1$ for the
$2p,3p \rightarrow 1s$ transition[3], $E2$ for the $3d \rightarrow 1s$[4,5],
and even $E3$ for the $3d \rightarrow 2p$ transition[6,7]
are all of approximately the same probability.
Experiments fully confirmed the Wheeler's conjecture and the further
predictions[8].

In ref.[6,7] special attention was brought to the fact that the radiationless
transition is a reverse conversion process. It was considered in terms of the
radiative nuclear width and internal conversion coefficients(ICC). That explained
the reason why higher multipole transitions turn out to be of about
the same probability. In 1958, Morita proposed a similar process of Nuclear
Excitation in Electron Transition, which is since known as NEET[9].
Calculations[6,7] revealed the strong resonance coupling arising in atoms
between the nucleus and the shell, or the muon in the case of muonic atom.
This coupling offers a real tool of mastering nuclear electromagnetic transitions
through resonance with the electron shell, using laser or synchrotron radiation.

The present paper is built as follows.
(i)We remind in main features general internal conversion(IC) theory.
(ii)We present the theory of the resonance internal conversion and outline
experiments where it was discovered.
(iii)We propose a way of mastering the rate of the resonance conversion by
applying external laser or synchrotron radiation, and consider concrete examples.  \\

\section{Internal conversion}

As a result of prompt fission, the muon is entrained on one of the fragments,
mostly on the heavy one. The fragment is excited, and emits
$\gamma^{\prime}s$, neutrons, protons, alphas.
Emitted $\gamma^{\prime}s$ can be absorbed by the muon, which leaves the atom (Fig.1).
\begin{figure}[!hbt]
\vspace{0.1cm}
\centerline{\psfig{figure=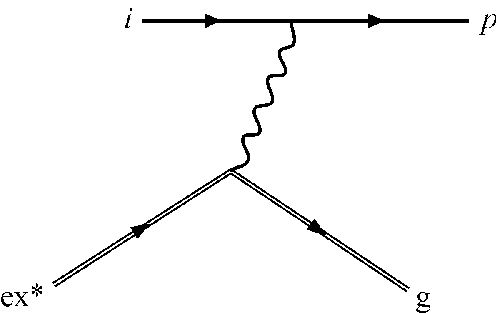,width=5.5cm}}
\vspace{0.1cm}
{\hsize=8.cm \leftskip=0.5cm \footnotesize{
{\bf Fig.1.}
Feynman graph of internal conversion.}
}
\label{find1}
\end{figure}
Such a process is well known in electronic atoms. It is called
\emph{internal conversion} (IC).
The necessary condition is
\begin{equation}
\omega_n>I
\end{equation}
where $\omega$ is the nuclear transition energy and I is the
ionization potential.

A very useful value is ICC, which is defined as the ratio of the conversion
and radiative transition probabilities,
\begin{equation}
\alpha(\tau L)=\frac{\Gamma_c}{\Gamma_{\gamma}^{(n)}}, \label{icc}
\end{equation}
$\tau$, $L$ being the type and multipole order of the transition.
$\alpha$ is nearly independent of the nuclear model.
This allows one to reverse eq.(2) and put down
\begin{equation}
\Gamma_c=\alpha\Gamma_{\gamma}^{(n)}.
\end{equation}
Values of $\alpha$ are tabulated. Eq. (3) hence allows one to
estimate the conversion transition probability,
as soon as the radiative nuclear width is known.
IC is one of the principal tools for nuclear spectroscopy.\\

\section{Resonance or Bound IC (BIC)}

If condition (2) is broken, i.e. $\omega_n<I$,
the conversion electron cannot leave the atom. It occupies an excited electron level,
forming an unstable intermediate state (Fig.2).
\begin{figure}[!hbt]
\vspace{0.1cm}
\centerline{\psfig{figure=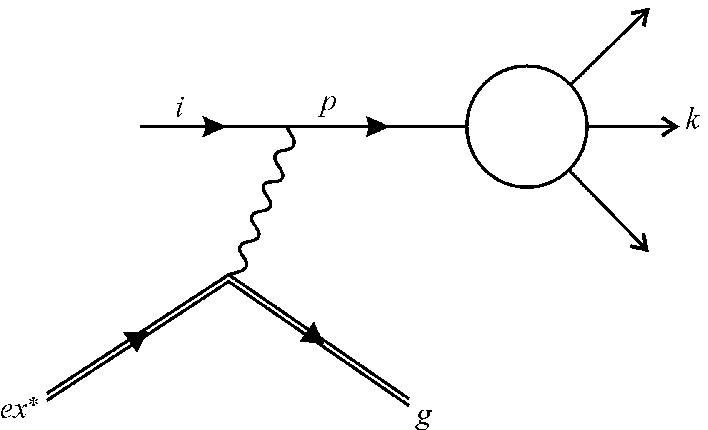,width=6.7cm}}
\vspace{0.1cm}
{\hsize=8.cm \leftskip=0.5cm \footnotesize{
{\bf Fig.2.}
Feynman graph of the resonance internal conversion.}
}
\label{rclf}
\end{figure}
This state then undergoes decay.  This is mainly performed by the
radiative electron transition,
filling the hole formed in the place of the conversion electron.
We still can formally calculate the $\alpha$ value by means of the formula,
used for the traditional ICC calculations,
inserting the wave function of the related discrete atomic state as
the conversion electron function:
\begin{eqnarray}
{\alpha}_d{({\tau} L)} = {\sum}_{\kappa}|M_{\kappa}^{({\tau}L)}|^2,
\hspace*{0.1em}
M_{\kappa}^{({\tau}L)} = Q_{\kappa}^{(L)} R_{\kappa}^{({\tau}L)} \label{kmkd2}
\end{eqnarray}
\begin{eqnarray}
Q_{\kappa}^{(L)} =
- \sqrt{\frac{\alpha \pi \omega}{L(L+1)}}C_{j_i-{\oh} L0}^{j_f -{\oh}};
\nonumber
\end{eqnarray}
\begin{eqnarray}
R_{\kappa}^{(ML)} = ({\kappa}_1+ {\kappa}) (R_1+R_2) \nonumber
\end{eqnarray}
\begin{eqnarray}
R_{\kappa}^{(EL)} = L (R_3+R_4)+({\kappa}_i- \kappa_f -L)R_5 + \nonumber \\ +
({\kappa}_i-\kappa_f +L)R_6 \nonumber
\end{eqnarray}
with $R_i$ --- the radial integrals
\begin{eqnarray}
R_1 = \int_{0}^{\infty} G_i F_f X_L dr & \hspace*{0.1em} &
R_2 = \int_{0}^{\infty} F_i G_f X_L dr \nonumber \\
R_3 = \int_{0}^{\infty} F_i F_f X_L dr & \hspace*{0.1em} &
R_4 = \int_{0}^{\infty} G_i G_f X_L dr \nonumber \\
R_5 = \int_{0}^{\infty} G_i F_f X_{L-1} dr & \hspace*{0.1em} &
R_6 = \int_{0}^{\infty} F_i G_f X_{L-1} dr. \nonumber
\end{eqnarray}
Here $\alpha_d{(\tau L)}$ is the discrete ICC, $\tau L$ is the type and
multipole order of the transition,
$EL$ and $ML$ stand for the electric and   for the magnetic types, respectively.
Subscripts $i$ and $f$  denote the initial and final states,
respectively;  $\kappa = (l-j)(2j+1)$ is the relativistic quantum number,
with  $l$, $j$ for the orbital and total angular moments;
$G$ and $F$ are the big and small components of the radial wave function,
 normalized at
\beq
\int_0^\infty [G^2 + F^2(r)] \,dr = 1 \;\;.
\eeq
Furthermore, $\alpha \approx$ 1/137 is the fine structure constant,
$M_{\kappa}^{({\tau}L)}$ are the conversion matrix elements, $\omega$
is the nuclear transition energy.
$R_{1-6}$ are the radial integrals, with
$X_\nu$ the interaction potential of the nuclear and electron transition
current.
With account of the finite nuclear size, the latter becomes
\beq
X_{\nu} = h_{\nu}(\omega r)
\label{kmkd3a}
\eeq
for $r  \geq R_0$, with $R_0$ --- the nuclear radius.
For $r < R_{0} $, nuclear model of the surface transition current
provides an adequate description\cite{sliv}:
\beq
X_{\nu} = \frac{h_{\nu}(\omega R_{0})}{j_{\nu}(\omega R_{0})}
j_{\nu}(\omega r).	\label{kmkd3b}
\eeq
In the case of discrete conversion,  however, the $\alpha_d$ value
acquires dimension of energy, due to another normalization of the
wavefunction\cite{mux}.
Therefore, it cannot serve as ICC (\ref{icc}) anymore.
There is an evident way, to form a dimensionless factor $R$ by
multiplying $\alpha_d$ by the resonance Breit-Wigner factor.
We add a subscript $d$ to the sign of $\alpha_d$, to distinguish it
from a traditional ICC. Then the expression for $R$ becomes as follows:
\begin{equation}
R=\frac{{\alpha_d\Gamma}/{2\pi}}{\Delta^2+(\Gamma/2)^2},
\end{equation}
where $\Gamma$ is the full width of the intermediate atomic state,
and $\Delta$ is the defect of the resonance of the nuclear and electron
transitions.
With the account of BIC, resulting lifetime of the nuclear level will be
\begin{displaymath}
\lambda=\frac{\lambda_\gamma}{1+\alpha_{tot}+R},
\end{displaymath}
where $\lambda_\gamma$ is the radiative lifetime, and $\alpha_{tot}$
is the total ICC.\\

\section{Tuning BIC}

It follows from eq.(4) that the BIC probability can be enhanced in the
case of resonance by the value of
\begin{equation}
\frac{R_{\rm res}}{R}\simeq\left(\frac{\Delta}{\Gamma}\right)^2   \;\;.
\end{equation}
For nuclei, typical values of $\Delta\simeq1{\rm keV}$.
Typical value of $\Gamma$ is $\sim$20~eV (which is a typical $K$-hole width),
or \mbox{$\sim10^{-6}$--$10^{-5}$~eV} in the case of BIC in the outer electron shells.
Therefore, expected effect can be around ten orders of magnitude and more
\cite{kabzon,atallah}.

The idea is to arrange a two-photon resonance. Consider atom in an external
field of a plane electromagnetic wave. Some atomic electron can go to an
excited state
by absorption of one or several photons from the field.
The probability of multiphoton absorption increases drastically
if the total energy of the absorbed photons approaches
the difference of the electron levels.
This effect is used by RIS --- resonance ionization spectroscopy.
Let us consider the two-photon resonance,
and replace one photon of the field by the nuclear photon \cite{HF}.
This is laser assisted nuclear BIC transition,
as the electron makes a two-photon transition to an excited state,
one photon being from the nucleus,
and the other from the field.
Necessary condition is then
\begin{equation}
\omega_n\pm\omega_l=\omega_a \;\;.
\end{equation}
The two signs in (6) correspond to either an absorption,
or induced emission of a photon of the field.
The both probabilities are of the same value.

\section{Nuclei in electromagnetic field}

Typical nuclear sizes are of the order of $R\simeq10$~Fm.
Typical scale for  their transition energies  is of $\sim$1~MeV.
Therefore, meaningful effect of the electromagnetic field on nuclei
can be only expected for fantastic electrical strength of the field
$\epsilon\simeq10^{18}V/cm$. At such strength, spontaneous generation
of $e^+e^-$ pairs already takes place (Klein's paradox\cite{klein}).
Such a simple estimate helps to realize that all the photo-nuclear
reactions are due to resonance effect. Only resonance quanta can be absorbed
by the nuclei, with the energy which exactly equals the nuclear level
separation energy.

Another lesson is why the only isomer was probably triggered up to now,
that of $^{180{\rm m}}{\rm Ta}$,
in spite of tremendous experimental efforts applied in this field[13].

Finally, atomic size is by four orders of magnitude larger than the
nuclear one. This means that it is much easier to affect the nucleus by
electromagnetic field through mediation of the electron shell,
which plays a role of resonator\cite{HF}.
In view of the above-mentioned difficulties related with triggering
the isomers, one must not afford a neglecting of several orders of
magnitude gain which can be benefited from making use of the resonance
properties of BIC. Further experimental study in the field must be
directed in this way.
We show in a separate paper  a concrete example of
this idea as applied to the $^{178{\rm m}2}{\rm Hf}$ isomer.  \\

\section{Resonance conversion in experiments}

A bright example of resonance effect of BIC is provided by characteristic muonic
X-rays from heavy prompt fission fragments.
This radiation arises due to resonance excitation of the bound muon to the $2p$ state,
with reemission in the succeding back muon transition $2p\rightarrow 1s$.
This resonance effect predicted in ref.[10]
is shown in Fig.3. The effect
\begin{figure}[!hbt]
\vspace{0.1cm}
\centerline{\psfig{figure=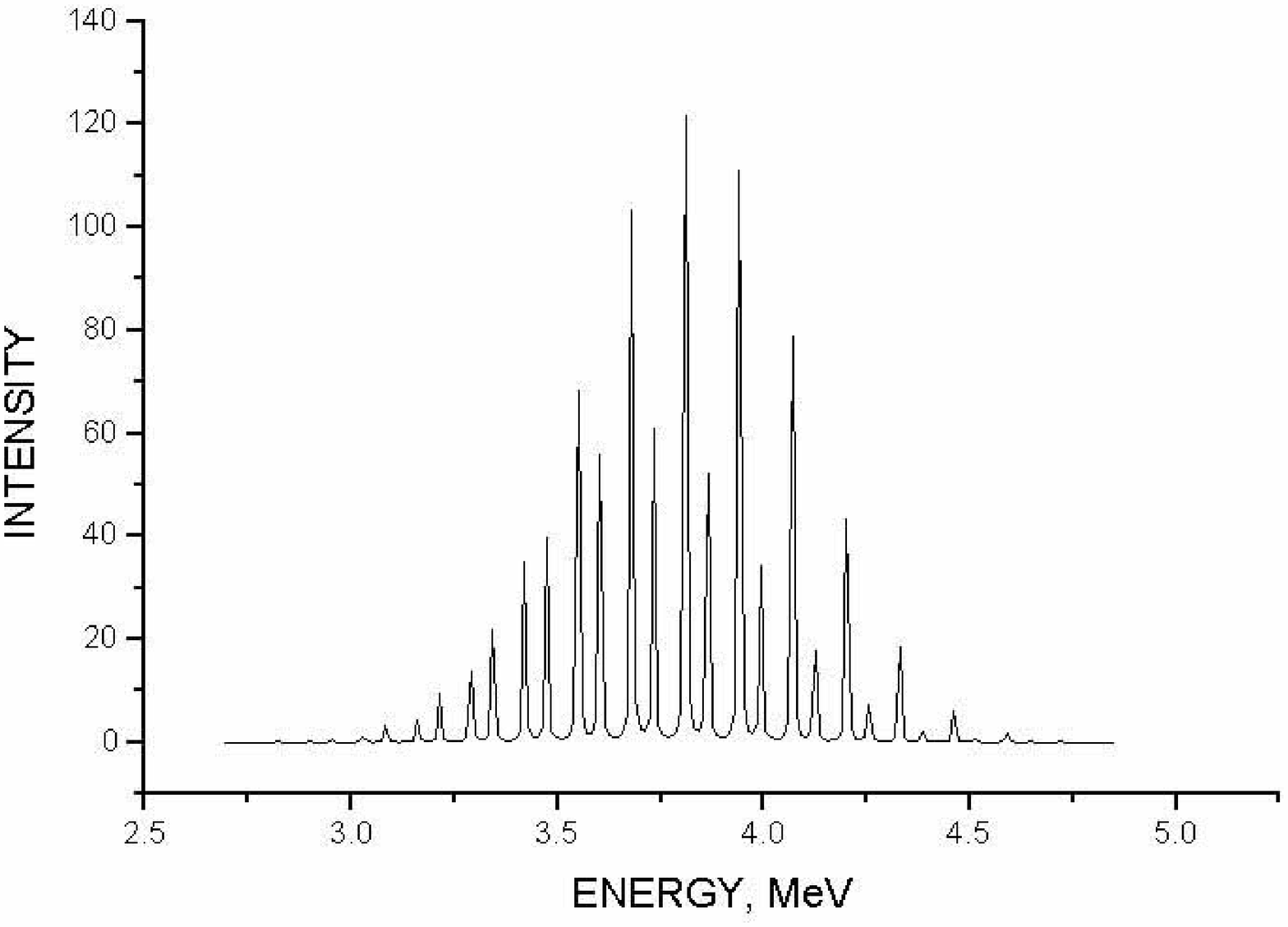,width=7.3cm}}
\vspace{0.1cm}
{\hsize=8.cm \leftskip=0.5cm \footnotesize{
{\bf Fig.3.}
Spectrum of $\gamma$ rays from prompt fission fragments, calculated with the account of the $\mu$ in the orbit.}
}
\label{muxf3}
\end{figure}
was experimentally studied in paper[15].
It was shown that taking this process into account leads to better value of
$\chi^2$
in filling the experimental spectrum of $\gamma$ rays from prompt
fission fragments.

The other example was demonstrated by the 35-keV $M1$ transition in
$^{125}{\rm Te}$\cite{atallah}.
In neutral atom, this level mainly deexcites via IC in the $K$ shell,
with \mbox{$\alpha(M1)$ $=11.5$}.
But for 45- and 46-fold ions the condition (1) is broken.
It was therefore expected that the lifetime will increase for these ions
by an order of magnitude (Fig. 4).
\begin{figure}[t]
\vspace{0.1cm}
\centerline{\psfig{figure=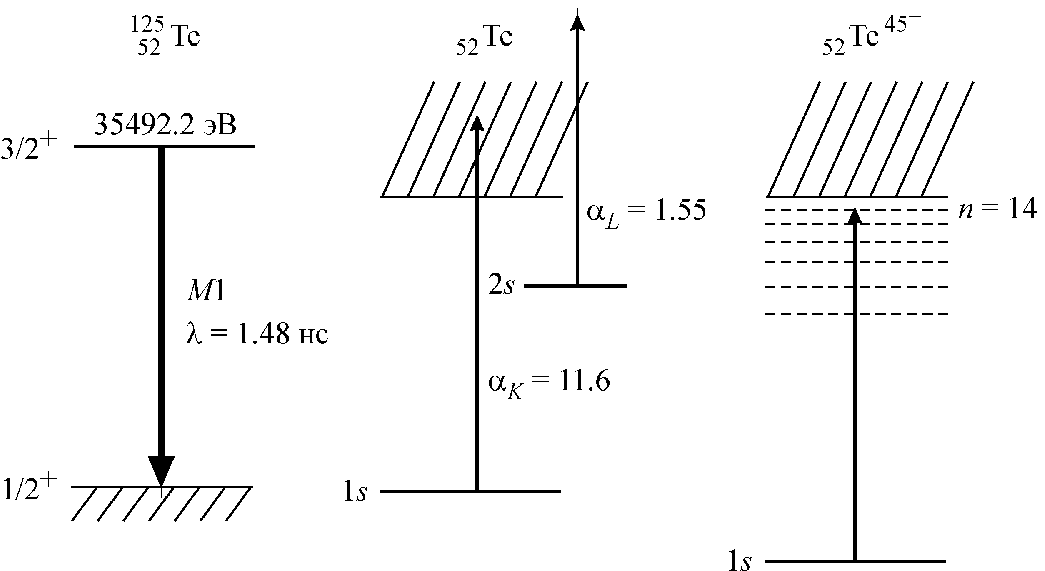,width=7.3cm}}
\vspace{0.1cm}
{\hsize=8.0cm \leftskip=0.5cm \footnotesize{
{\bf Fig.4.}
Scheme illustrating how traditional internal conversion transforms into
the resonance one in ionized atoms of $^{125}$Te.}
}
\end{figure}
Experiments however showed that the lifetime holds\cite{atallah}.
The observed paradox was explained by the resonance conversion --- BIC,
which takes place of the ordinary one.
In the succeding papers, the fluorescence yield from the $1s$-hole states
caused by the resonance conversion was observed.   \emph{The observed
numerical probability of BIC came to  accordance with theory.}

A very promising experiment can be undertaken in hydrogen-like
ions of $^{169}$Yb. The energy of the $M1$ 70.6-keV transition differs
by about 10~eV from the energy of the atomic $1s \to 10s$ transition.
Therefore, it is expected that the nuclear lifetime shall be
considerably shortened in the field of electromagnetic wave of
such frequency\cite{Yb}. This effect is demonstrated in Fig.5.
As one can see, the effect can achieve up to four oders of magnitude.

A very promising field of application of the resonance BIC is the laser
produced plasma. Then NEET, which is a reverse BIC process,
is one of the possible mechanisms which can lead to formation of nuclear
isomers  in a heat bath\cite{ter2,inam}.
This mechanism is under experimental investigation at the time being,
specifically, with respect to the 76-eV $^{235{\rm m}}{\rm U}$
isomer\cite{Brd235},
1.5-keV $^{201}{\rm Hg}$ isomer\cite{Meot},
6-keV $^{178}{\rm Ta}$ isomer \cite{Andreev}, and others.
Further peculiarities of NEET arising in plasma are considered in
paper\cite{poz01}.

Some of the results presented above were delivered on the conference
\cite{fdacn} held under honorary patronage of J. Wheeler.
He was satisfied to know about development of his idea.
\begin{figure}[!h]
\vspace{0.1cm}
\centerline{\psfig{figure=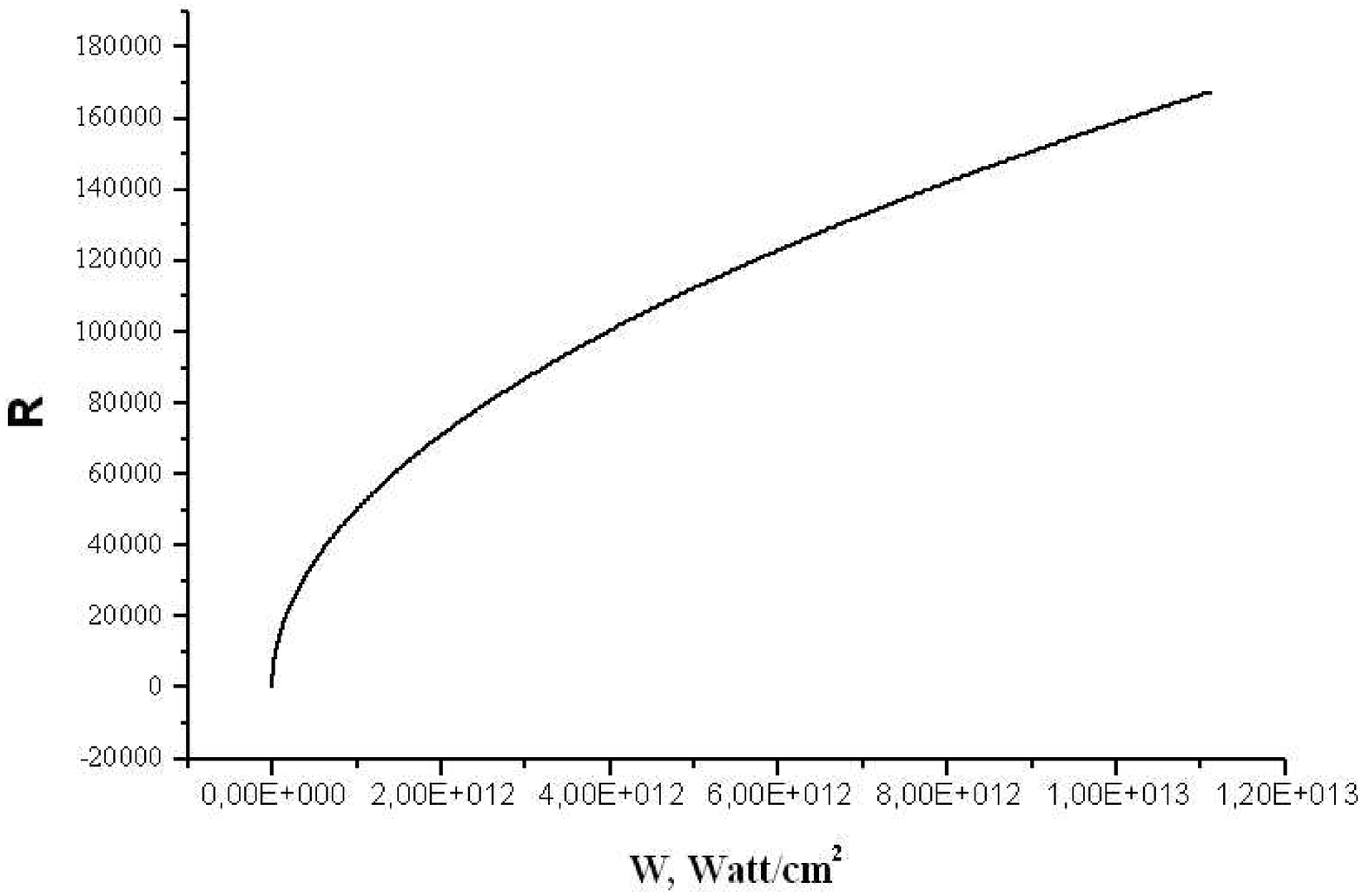,width=7.3cm}}
\vspace{0.1cm}
{\hsize=8.cm \leftskip=0.5cm \footnotesize{
{\bf Fig. 5.}
Values of the resonance enhancement factor $R$ calculated for the hydrogenlike ions of $^{169}$Yb versus intensity of the external field.}
}
\end{figure}

\end{document}